# Non-utopian optical properties computed of a tomographically reconstructed real photonic nanostructure

Lars J. Corbijn van Willenswaard,[1,2] Stef Smeets,[3] Nicolas Renaud,[3] Matthias Schlottbom,[2] Jaap J. W. van der Vegt,[2] and Willem L. Vos[1,*]

[1]*Complex Photonic Systems (COPS), MESA+ Institute for Nanotechnology, University of Twente, P.O. Box 217, 7500AE Enschede, The Netherlands*
[2]*Mathematics of Computational Science (MACS), MESA+ Institute for Nanotechnology, University of Twente, P.O. Box 217, 7500AE Enschede, The Netherlands*
[3]*Netherlands eScience Center, Science Park 402, 1098 XH Amsterdam, The Netherlands*
https://nano-cops.com
[*]*w.l.vos@utwente.nl*

**Abstract:** State-of-the-art computational methods combined with common idealized structural models provide an incomplete understanding of experimental observations on real nanostructures, since manufacturing introduces unavoidable deviations from the design. We propose to close this knowledge gap by using the *real structure* of a manufactured nanostructure as input in computations to obtain a realistic comparison with measurements on the same nanostructure. We demonstrate this approach on the structure of a real inverse woodpile photonic bandgap crystal made from silicon, as previously obtained by synchrotron X-ray imaging. A 2D part of the dataset is selected and processed into a computational mesh suitable for a Discontinuous Galerkin Finite Element Method (DGFEM) to compute broadband optical transmission. We compare this to the transmission of a utopian crystal: a hypothetical model crystal with the same filling fraction where all pores are taken to be identical and circular. The shapes of the nanopores in the real crystal differ in a complex way from utopian pores due to scallops, tapering, or roughness. Hence, the transmission spectrum is complex with significant frequency speckle both outside and inside the main gap. The utopian model provides only limited understanding of the spectrum: while it accurately predicts low frequency finite-size fringes and the lower band edge, the upper band edge is off, it completely misses the presence of speckle, the domination of speckle above the gap, and possible Anderson localized states in the gap. Moreover, unlike experiments where one can only probe from the outside of a real crystal, the use of a numerical method allows us to study all fields everywhere. While at low frequencies the effect of the pore shapes is minimal on the fields, major differences occur at higher frequencies including the gap such as high-field states localized deep inside the real crystal. We conclude that using only external measurements and utopian models may give an erroneous picture of the fields and the local density of states (LDOS) inside a real crystal, while this is remedied by our new approach.

## 1. Introduction

The field of nanophotonics promises control over emission and propagation of light by tailoring the environment [1–3]. Photonic crystals are particularly promising tools for the goal of simultaneous control of light in all three dimensions (3D). Such photonic crystals are defined by a periodic variation of the refractive index commensurate with optical wavelengths, which results in photonic dispersion relations organized in bands, analogous to the electron bands in solids [4,5]. Analogous to the electronic band gap, there is a complete 3D photonic band gap, a frequency gap without any states. Within the 3D band gap, no light modes exist inside the crystal, hence the density of states (DOS) strictly vanishes. In finite crystals, the local density of states does not vanish completely, but by making the crystal large enough it can be suppressed to arbitrarily low levels [6–8]. A 3D gap is therefore a powerful tool to radically control the spontaneous emission and cavity quantum electrodynamics (QED) of embedded quantum emitters [9,10]. There is a wide range of applications of 3D photonic band gap crystals, including efficient photovoltaic cells [11,12], miniature lasers [13], thermal emission control [14,15], mode and polarization converters [16], cloaking devices [17], and 3D photonic integrated circuits [18,19].

Inevitably, every fabricated 3D photonic crystal will, like any nanostructure, deviate from its design, both systematically in the form of structural deformations and statistically in shape and size variations. Consequently, the optical properties of actual samples differ from those of the design [20–22]. Testing the optical properties of a real sample, for example by measuring a stopband in reflection, typically shows a complex signal that differs markedly from theoretical predictions [23]. These differences could not only be caused by the manufacturing deviations, but also by the experiment or the model that was used to predict the optical properties in the first place. To guide design, experiment, and modeling of nanophotonic devices, it is thus crucial to understand how real devices with manufacturing deviations differ from the perfect devices that could only be manufactured in a utopian cleanroom. Here, we propose and demonstrate a completely new approach to nanophotonics to bridge the gap between model and experiment. We expect that our approach will boost the application potential of numerous classes of nanophotonic platforms, not only 3D photonic band gap crystals, but also systems with great complexity such as metamaterials [24], stacks of chiral materials [25], nanoparticles [26,27], waveguides [28], non periodic structures [29], as well as the robustness of 3D topological photonic materials to unavoidable disorder [30].

## 2. Interpreting experiments with models

### 2.1. Traditional approach

To put our new approach in perspective to the current traditional approach to nanophotonics, it requires that we understand the limitations of the current approach. As with most physics research the traditional approach consists of two tracks: experiments and theory. The central idea is that if we can explain a set of experiments with a theory then we understand what is happening. In reality, this is quite difficult with large nanophotonic devices, because there are significant complications and limitations on both the experimental and the theoretical tracks.

To understand the limitations of the traditional approach, illustrated in Fig. 1(a), let us consider having a design for a large nanophotonic device, *e.g.*, a 3D photonic band gap crystal, whose physical properties we seek to understand. The experimental route consists of two steps: The design is used to manufacture a number of samples, a beam with photonic crystals in our case. Each sample is then mounted into an experimental setup so that its optical properties are measured, for example, the reflectivity as function of frequency [23,31]. The theoretical route consists of putting a model of the design into a computer for a computation using a numerical method like a Finite Element Method (FEM) [32], Plane Wave Expansion (PWE) [33], or Finite Difference Time Domain (FDTD) method [34]. Which optical properties can be computed in this way depends on the model, but in theory they are expected to match the experimental results. In practice this is much more difficult, and with a complicated system like a photonic crystal one is happy with a limited match between theory and experiment, for example, the center frequency of a measured reflectivity peak matches the center of a theoretically predicted bandgap [35,36].

The sources of these complications lie in all three steps, (1) computation, (2) experiment and (3) manufacturing. We discuss the complications introduced by each of these steps individually.

(a) Traditional

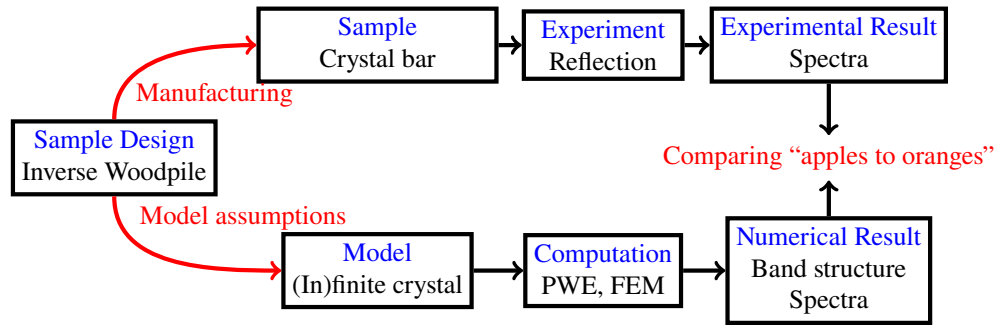

(b) Data-driven

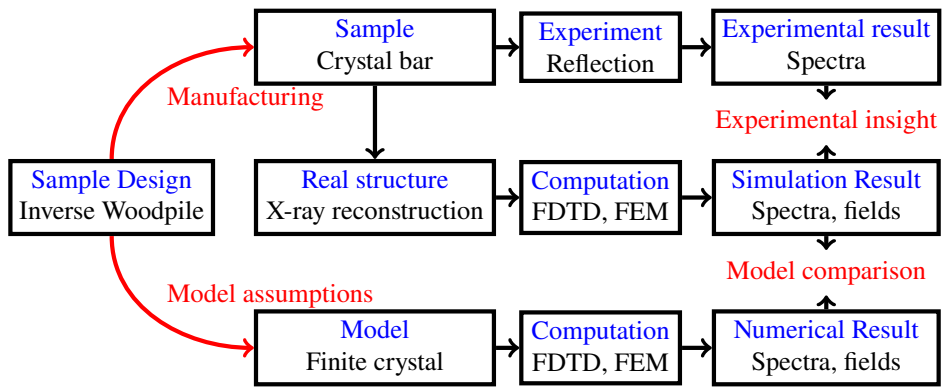

**Fig. 1.** Schematics of the traditional approach to nanophotonics research and our new data-driven approach. (a) The traditional approach compares numerical results based on an idealized utopian model of the design sample with experimental results of a real manufactured sample. Each conceptual step (blue) is illustrated with an example (black). Both the real sample and numerical models differ from the original design, the former due to fabrication, the latter due to modeling assumptions (red arrows). These differences fundamentally limit the comparison between the numerical and experimental results. (b) The data-driven approach using the real structure as input for a simulation can mediate between the experimental and utopian model results.

(1) Computationally the challenges in essence revolve around the finite computer memory and computation time. A realistic simulation of the experimental situation is beyond the ability of current (super)computers for anything but the tiniest photonic crystals. Therefore, the computational approach has to make modelling assumptions and simplifications. Typical examples are simplifying an incident beam from a focus to a plane wave, using an infinite rather than a finite crystal, or using only simple geometric shapes rather than a freeform structure. These assumptions affect the computed optical properties and thus limit and complicate a comparison with experiment. An extreme example of this is the use of an infinite crystal and the computation of a band structure. The lack of any external interface and of finite size effects means that such a band structure is limited in what it can predict and explain about experiments.

(2) Challenges with the experiment are primarily practical, *i.e.*, we need to get a reliable signal for the optical property that we want to measure. The extent and design of photonic crystals means that we are limited in what kind of signals we can obtain, as we can not "just stick a probe in there", especially not in 3D nanostructures [37]. Even if we could, this probe would probably affect the optical properties under study. Moreover, not all effects give a significant signal, for example, measuring the effect of a bandgap through transmission is hard because the transmission in a bandgap should vanish, but a vanishing signal could also be caused by a myriad of other reasons, and such a signal is thus potentially unobservable due to noise.

(3) The central challenge with manufacturing is that it will inevitably introduce deviations from the original design, thereby altering the optical properties [22,38,39]. The current techniques for manufacturing photonic crystals will leave artifacts, and these artifacts are large enough that they significantly impact device performance, as will be shown in our results. These artifacts can roughly be grouped into two categories, systematic and random. A systematic deviation is one where the whole crystal deviates from the design in a predictable manner, for example, the misalignment of two parts of the design [20,40]. With random deviations the actual crystal deviates from the design in a way that can only be characterized statistically, for example, if the shape of each feature is deformed in a different way.

The current traditional approach to estimate and understand the effects of manufacturing deviations is to use computational and theoretical models. This is very useful for systematic manufacturing deviations, as one can typically model these with one or a few parameters, like an offset parameter to model a misalignment. Studying the effect of such a parameter is the same as optimizing any other design parameter like lattice constants and feature sizes [40]. There are three problems with this approach for predicting the effect of random manufacturing deviations. Firstly, one needs to have a parameterized model to generate such deviations. But deviations like shape variations are very difficult to realistically model with a tractable number of parameters. Secondly, the randomness means that one needs a realistic probability distribution for the generation of these deviations, but there is typically no information about these distributions in a real crystal, with a few rare exceptions [41]. Thirdly, the randomness means that the optical properties are only valid for one specific generated realization, which is not necessarily realistic [20]. One could work around this by using ensemble averaged properties [42]. However, such statistical quantities only give information about an ensemble of samples and are therefore unsuitable to explain what happens in a single sample with a specific configuration [22,43].

### 2.2. Our data-driven approach

Our new data-driven approach tackles the problem of the unrealistic utopian structure used in the computations, as illustrated in Fig. 1(b). We do this by extracting the crystal structure from a tomography reconstruction of a *real crystal, thereby including all manufacturing details*. This structure is then used for a computation of the optical properties of the photonic crystal. We envision two ways how our new approach can complement the current approaches.

(1) The comparison between measurements from a real crystal with the computation that is based on the structure of the same crystal. By using the real structure of the crystal one includes the exact manufacturing deviations from the utopian design, just as in the experiment. The effect of these deviations should therefore be the same in both experiment and computation thereby obviating many assumptions, giving a more accurate comparison. Moreover, as we can study the field inside a crystal with a computation, we can effectively investigate effects that are difficult or even impossible to probe in an experiment.

(2) The comparison between computations with a real structure and computations with related utopian model geometries can be used to determine how accurate the model is as predictor for a real nanostructure. Optical properties that are both present in the model and the real-structure computation are likely relatively robust with respect to manufacturing deviations and thus also present in a real nanostructure. Properties of the model that are absent in the real-structure computation are likely sensitive to manufacturing deviations and thus not present in the real nanostructure.

The use of the structure of a real nanostructure in our approach is what makes these applications possible and what makes it fundamentally different from previous approaches with modeled manufacturing deviations. These previous modeled manufacturing deviations are limited by human imagination, which includes parametrizing them and finding a proper distribution for the randomness. The use of the structure of a real sample bypasses these problems as it uses the actual physical process to generate a real computational version of manufacturing deviations according to the actual statistical distribution. Moreover, because the reconstruction is of a real nanostructure, we are not limited to ensemble averaged quantities, as the actual computed properties are relevant for that specific sample, and in terms of statistical physics: on that specific configuration [43].

We envision the steps shown in Fig. 1 to be a part of a larger cyclical design process, where results as obtained here provide advanced feedback to considerably improve the design, manufacturing and modeling of a nanostructure. To understand how our new data-driven approach improves this cyclical process, we consider its three main requirements: Firstly, our approach needs a manufactured nanostructure of sufficient quality, secondly, a process to scan this sample with sufficient detail to reveal manufacturing deviations in the actual sample, and thirdly, sufficient computational resources to simulate the scanned nanostructure (even if only in part) including the scanned manufacturing details. This points to a design stage where the user has a good starting design and is interested in verifying, understanding, and optimizing an existing design and manufacturing process. Modern techniques for this design stage already include 3D (X-ray) imaging to verify the internal structure, and large computational resources for topological optimization of the structure, thus making our new data-driven approach viable for a wide range of photonic nanostructures.

## 3. Methods

### 3.1. Design and dataset

The unit cell design for the inverse woodpile photonic crystal structure is shown in Fig. 2(a)–2(c) [44]. The crystal structure consists of intersecting pores with radius $r$ along the $X$ and $Z$ direction. The pores in the $X$ and $Z$ directions form a centered rectangular lattice in the $YZ$ and $XY$ planes, respectively. This crystal has lattice parameters $a$ and $c$, with $c = a/\sqrt{2}$, such that the crystal has a face centred cubic (fcc) lattice, with a diamond like symmetry [44,45].

In practice, real 3D photonic bandgap crystals are manufactured from silicon on the edge of a beam using methods described in Refs. [46–48]. In brief, a mask with an array of apertures is made on both the top and front sides of the beam, followed by two consecutive reactive ion etching steps to etch nanopores both from the top and the side of the Silicon beam. The volume where the two pore sets overlap is a 3D periodic nanostructure with the inverse woodpile structure.

A number of these crystals were non-destructively imaged using X-ray imaging (holotomography) [36,47,49–51]. Tomographic reconstruction was used to obtain the electron density inside the scanned volume. The resulting datasets are divided in cubic voxels, with each voxel associated with the electron density in its volume. We note that obtaining quantitative electron density is a major step forward over the arbitrary units typical of scanning electron microscopy (SEM). From all datasets we choose the one with the smallest voxels and the largest contrast in electron density, expecting this will give the most accurate reconstruction. This specific dataset has $1024 \times 2048 \times 1024$ cubic voxels with 10 nm edges. The axes of this dataset are aligned to be parallel to the top and side plane of the beam with crystals. Moreover, since the pores extend perpendicular from the surface of the beam, this alignment results in pores that are also aligned with the dataset axes.

Figure 2(d) shows part of the dataset with size $800 \times 1108 \times 800$ voxels that corresponds to 33% of the whole dataset. It highlights a single crystal on the silicon beam. The variations on the three image planes correspond to the intersection with the crystal pores. Periodic variations are apparent from intersections with the 2D lattices of pores. The lengthwise cross-section is visible on the *XZ* plane, which also shows that the pores in the *Z* direction are less than a third of the length and thinner than those in the *X* direction.

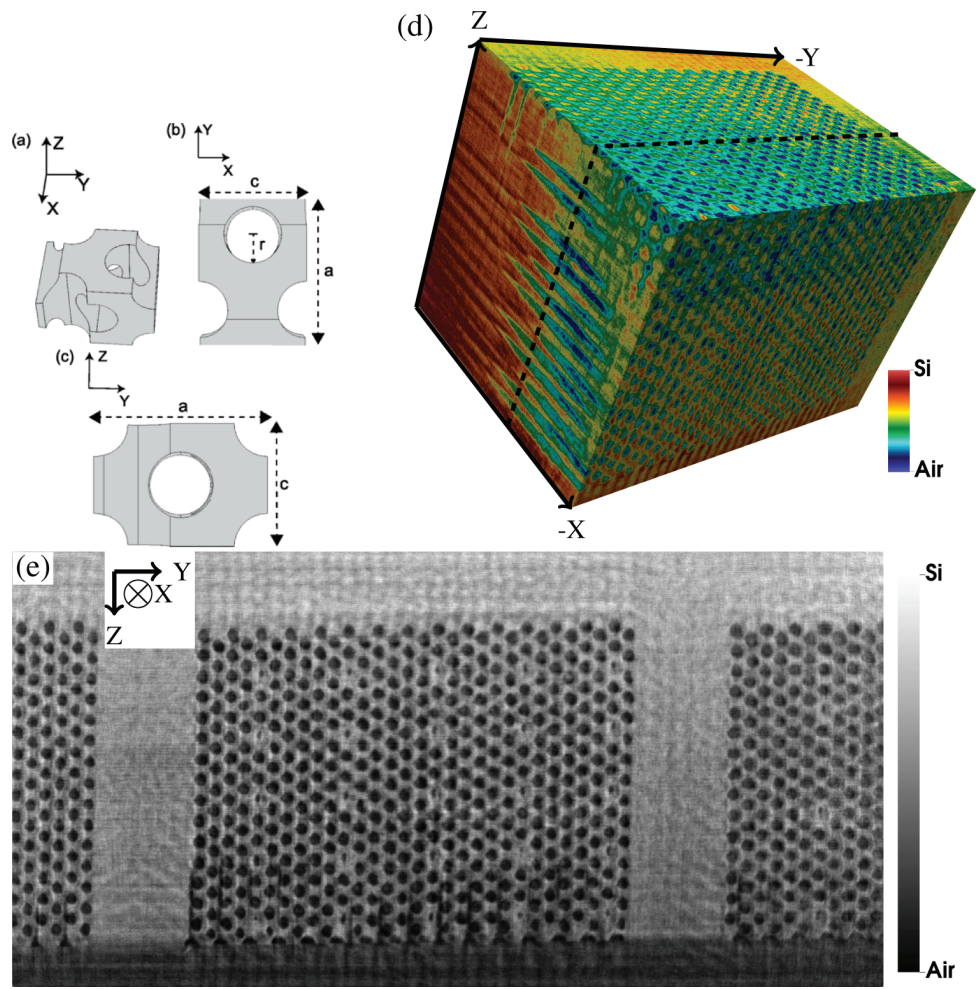

**Fig. 2.** (a,b,c) Schematic of the inverse woodpile unit cell, used as a basis for the design of the fabricated nanostructures. (d) Bird's-eye view of the X-ray holotomography dataset cut out to highlight a single crystal. (e) YZ-slice of the X-ray holotomography dataset (size 1024 × 2048 voxels), lighter regions have a higher electron density. This slice intersects the birds-eye view (d) at the dashed plane.

Figure 2(e) shows a complete slice of the dataset in the *YZ*-plane at the edge of the silicon beam, approximately 4000 nm below the top surface. Three photonic crystals are visible, a complete one in the middle and two partial ones on the left and right. The transition between dark area at the bottom (vacuum) and lighter area is the front surface of the silicon beam.

The lattice of circular dark areas correspond to the intersection of this slice with the pores in the *X*-direction. The periodic lighter pores in the middle and right crystal are pores with a reduced size in the design thereby creating a lattice of defects. The pores in the −*Z*-direction are short and thin and as result barely visible.

From this slice we estimate that the deepest pores in the −Z-direction are approximately 4 lattice constants deep, corresponding to approximately 1900 nm. Thus there are only pores in both the *X* and −*Z*-directions in the volume near the *XY*-surface. The crystal volume that is deeper only has pores in the *X*-direction, thus forming a quasi-2D structure, see also Ref. [47].

### 3.2. Data processing

Numerical computations on the total dataset volume with all $2 \times 10^9$ voxels are at this time not feasible as they require prohibitively large computational resources. As a result it would, if at all possible, put such severe restrictions on spatial and frequency resolution that a comparison to experiment would not be meaningful. Therefore, for this study we limit ourselves to a part of the 2D-slice of Fig. 2(e). The reason for choosing a part of a 2D-slice is that this reduces computational cost and allows high-resolution computations to accurately compute the field around the irregular pore shapes. While the 2D slice is not directly comparable to experiment, it does contain real manufacturing deviations and is thus interesting for quantifying the optical effects these manufacturing deviations have. The selected part is shown in Fig. 3(a), which is the largest region in the dataset that is in the 2D part of a crystal with uniform pores in the design. The edge of the region is chosen to cut through the middle of pores to allow mirroring the domain along the edges. After this mirroring the pores on the edge of the domain have nearly the same size as the pores that are completely inside the domain.

The processing of the dataset to a mesh for a finite element computation consists of three steps: Cutting out the region, extracting the pore-silicon interface, and meshing. These three steps are automated using Nanomesh [52], an open source Python library for processing 2D and 3D data into computational meshes. The code for all these steps and the actual computation are available as part of the dataset in Ref. [53]

The slice in Fig. 3(a) and the dataset in general shows high frequency noise and line artifacts. To reduce these artifacts the image is smoothed using a Gaussian filter with a standard deviation $\sigma = 30$ nm corresponding to 3 pixels. This agrees well with the 2.5 pixels spatial accuracy as determined in Ref. [36]. The smoothed image is segmented into two regions, air and silicon, using local thresholding [54,55]. The region is determined by comparing each pixel to an offset plus a weighted mean of the surrounding square of $101^2$ pixels. This use of a local threshold is essential because there is an intensity gradient in the dataset. With a global threshold this gradient would introduce a gradient in the pore sizes. The size of the comparison region and the threshold were visually chosen to result in pore outlines that conform to the X-ray dataset. Lastly, the marching cubes algorithm [56] is used to convert the segmentation into a pore boundary.

Figure 3(b) shows the silicon-air interface of the pores overlayed on the dataset after applying the Gaussian filter. Based on the inverse woodpile design [44] of this sample we had expected the pores to be circles or ellipses. However, both the reconstructed contours and the X-ray data of Fig. 3(a) shows that the pore shape and size varies significantly in the real crystal. Possible reasons for these deviations include unknown subtleties in the etching [48,57], and image artifacts in the X-ray holotomography reconstruction [36]. Nevertheless, the resulting pore shapes have similar shapes as those seen in SEM images of the external surface of the same crystal, indicating that the reconstruction is realistic.

The interfaces of the partial pores on the left and right boundaries are discarded to create a flat crystal interface, similar to the interface of the actual inverse woodpile. The interface from the segmentation step is used to constrain the meshing algorithm to get a mesh that exactly follows this interface. An example of such a mesh is shown in Fig. 3(c), which is the coarsest mesh used to determine the accuracy of our results (see Supplement 1). The transmission spectrum and fields presented here use triangles with about 4× smaller edges, thus yielding accurate results.

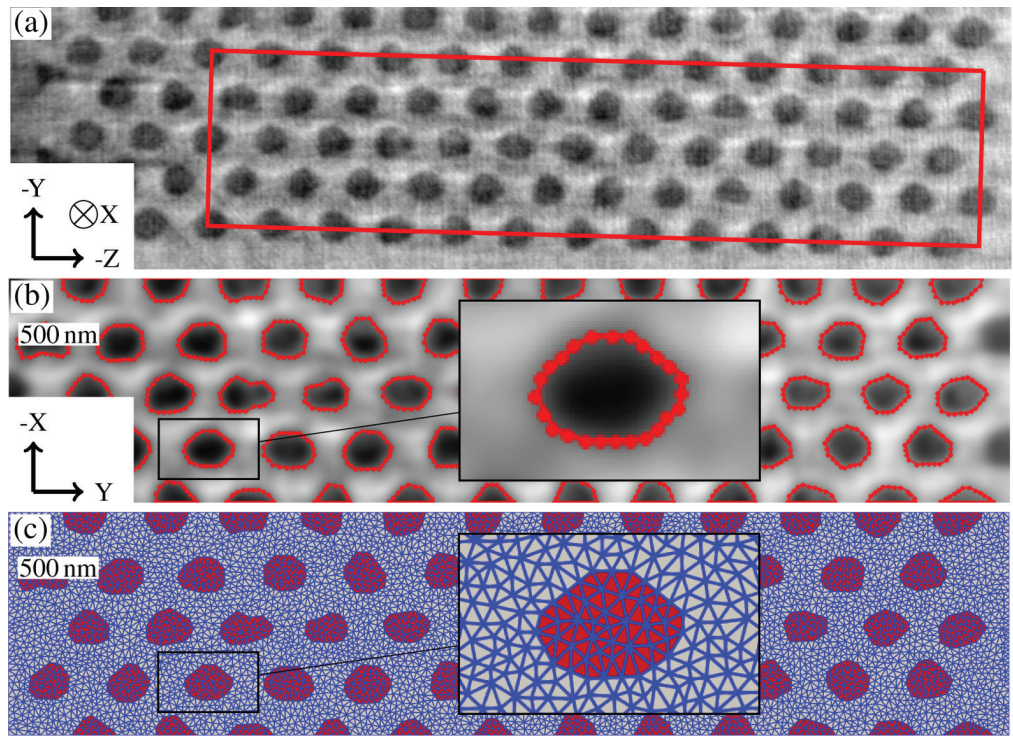

**Fig. 3.** Processing of the X-ray holotomography dataset to numerical input. (a) A small part of the electron-density dataset, light regions correspond to silicon, dark ones to air. The rectangular region outlined in red (size 6.19 µm × 1.37 µm) is selected for further processing. (b) A Gaussian filter is applied to reduce the high frequency noise in the electron-density, the result is segmented using local thresholding. The red lines are the border between the silicon matrix and the pores. The scale bar is 500 nm. (c) A mesh is generated for the whole region, with the segmentation border as constraint. The color of the elements corresponds to the material with in grey silicon, and in red the pores. Pores intersecting with the left and right boundaries have been removed.

### 3.3. Numerical

To compute the optical transmission of the reconstructed crystal we use a Discontinuous Galerkin Finite Element Method (DGFEM) to solve the time-harmonic Maxwell's equations [32,58–61]. The details of the method can be found in the Supplement 1, in short we use an Interior Penalty discretization for the time harmonic Maxwell's equations using piecewise second order Nédélec elements. Combined with the mesh that resolves the interface we accurately handle the sharp contrast in the dielectric constant between silicon and air.

The computational setup is shown in Fig. 4(a). The mesh obtained from X-ray image processing is extended by a layer of air on the left and silicon to the right. This is similar to the real crystal, which is on the edge of a silicon beam [46]. We take the relative permittivity of silicon as $\epsilon = 12.1$, matching with previous results [40,62]. On the top and bottom we use perfectly conducting boundary conditions to create mirror symmetry. By using this on both boundaries we obtain half of a structure that extends periodically in the vertical direction. Alternatively, the two perfectly conducting boundaries can be interpreted as making the domain into a strip waveguide, the resulting effects are further discussed in section 5.2

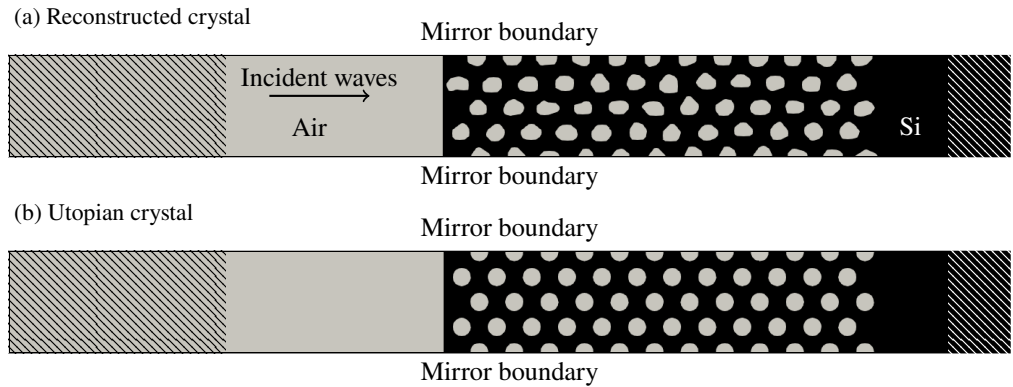

**Fig. 4.** (a) The setup for transmission computations. To mimic real crystals the reconstructed crystal structure is extended with air to the left and silicon to the right. Plane waves are send in from the left. The hatched regions on the left and right are PMLs which absorb the reflection and transmission from the crystal. For the top and bottom we use mirror boundary conditions (perfect electric conductor), which extend the domain periodically in the vertical direction. (b) The computational domain for the derived disorder-free utopian crystal.

We consider light with the magnetic field orthogonal to the computational plane, in photonic crystals known as TE polarization [1], and in waveguides as TM polarization [63]. As incident fields we send in a plane wave from the air region, with the electric field parallel to the air-crystal interface. This is the lowest order mode supported by the domain, denoted as the 0-th mode.

For the computation of the total electric field we use a scattering approach, where we consider the computational setup as an air-silicon interface with the air-pores as added scatterers. The total electric field is thus split into two parts $\mathbf{E}_{tot} = \mathbf{E}_b + \mathbf{E}_s$, a background field $\mathbf{E}_b$ from the interface and a field $\mathbf{E}_s$ from the scattering by the pores. The background field $\mathbf{E}_b$ consists of the incident plane wave and its reflection and transmission from the air-silicon interface, as computed using the Fresnel equations [63]. This background field is scattered by the pores, the resulting scattered field $\mathbf{E}_s$ is computed numerically. We use a Perfectly Matched Layer (PML) to prevent the scattered field from reflecting off the left and right boundary of the computational domain [64,65], the details of the PML can be found in the Supplement 1.

For the transmission we calculate the energy flux into the PML behind the crystal. We compute the transmittance by numerically integrating the normal component of the Poynting vector of the total electric field over the interface between silicon and the PML. To obtain the transmission we take the ratio of the transmittance to the energy flux from the incident plane wave, before its reflection from the air-silicon interface.

### 3.4. Utopian model crystal

The pores of the reconstructed crystal vary in shape and size from the perfect cylinders in the original design. The optical properties of the real crystal and our derived computational model will therefore differ from predictions based on the original design. To asses these differences we use the computational setup in Fig. 4(b). This setup uses a crystal structure with circular pores with the design parameters fitted to get optical properties closely matching to the reconstructed crystal. The structure can thus be seen in two ways, namely the structure that would have been made in a utopian world with perfect manufacturing, or the best predictive model that would have been used before manufacturing.

The utopian model is made by replacing the varied pores by circular pores placed on a 2D centered rectangular lattice. This is the same type of structure as is used for the etch mask.

This design has three parameters, the lattice constants $(a, c)$ and the pore radius $r$. These parameters were adjusted to match the geometric and optical design of the reconstructed crystal. We take the lattice parameter $a = 685$ nm to match the width of the reconstructed and design crystals. For the lattice parameter $c$ we have a choice, to measure it based on the sample length $c = 6.19\,\mu\text{m}/12.5 = 495$ nm or to reuse the aspect ratio of all our designs $c = a/\sqrt{2} \approx 484$ nm. We chose to keep the design aspect ratio, as this would have been used to predict the properties of the real structure before manufacturing. The pore radius $r = 126$ nm was chosen so that the model crystal has the same average pore area as the reconstructed crystal, and therefore the same average refractive index [66], and hence the same gap center frequency [67]. The triangles in the mesh have similar size to those in the mesh for the reconstructed crystal, ensuring similar accuracy of the computation.

## 4. Results

### 4.1. Transmission spectrum of the reconstructed crystal

Figure 5(a) shows the transmission spectrum of the reconstructed crystal between 2000 and 8000 cm$^{-1}$, in steps of 2 cm$^{-1}$. Between 4812 and 5600 cm$^{-1}$ the transmission is below $T<1\%$, indicating a photonic stopband. We distinguish three ranges in the spectrum, (1) below, (2) in and (3) above the stopband, each with different behavior, as discussed here. A log-transmission version of Fig. 5(a) is presented in the Supplement 1, as well as a convergence study demonstrating that the computational error in the wavenumber is less than 2 cm$^{-1}$.

Below the stopband the transmission shows broad fringes with some regions with sharp peaks. In this spectral range we distinguish three subregions, (1a) to (1c), depending on whether the fringes or peaks dominate the spectrum.

(1a) Below 3600 cm$^{-1}$ the fringes dominate the spectrum with only one small extra peak at 2790 cm$^{-1}$. The fringes have a peak-to-peak spacing of $\Delta \nu = 270$ to 300 cm$^{-1}$, and vary in amplitude from 17% to 39%. These fringes are typical for low frequencies, where a photonic crystal behaves as an effective medium [66]. If a crystal has flat and plane parallel front and back surfaces, the crystal behaves similar to a Fabry-Pérot etalon [68], revealing interference fringes as discussed in Refs. [35,62].

(1b) Between 3600 cm$^{-1}$ and 4550 cm$^{-1}$ the spectrum shows both fringes and sharp peaks. The amplitude and width of the fringes between 4200 and 4550 cm$^{-1}$ is comparable to the fringes below 3600 cm$^{-1}$. The sharp peaks in this regime have an amplitude of at most 30%, similar to the fringe amplitude. The onset of this regime is near 3649 cm$^{-1}$, where the second air mode of the computational domain transitions from evanescent to propagating, as discussed in section 5.2.

(1c) Between 4550 cm$^{-1}$ and the edge of the stopband at 4812 cm$^{-1}$ the spectrum is dominated by sharp peaks with amplitudes varying to more than 80%. A region of fast changes in transmission just below the stopband is similar to the compressed fringes below the stopband of a 1D layered photonic crystal [35,69]. Since the peaks reveal varying mutual spacing and bandwidth, it is also conceivable that they are in part speckle due to the randomness of the nanostructure [70].

(2) At the low frequency edge of the stopband, just below 4812 cm$^{-1}$ we see a sharp decrease from a bright $T>65\%$ transmission peak at 4794 cm$^{-1}$ to near zero transmission. Between 4812 and 5600 cm$^{-1}$ the transmission is less than $T<1\%$, hence a stopband of the crystal. This stopband matches well with the $\Gamma - K - B$ high symmetry stopgap predicted from 4771 to 5977 cm$^{-1}$ from Ref. [23], especially considering that they use a slightly higher $\epsilon = 12.2$. Just above the stopband, up to about 6000 cm$^{-1}$, the transmission increases with several very sharp peaks. The width of these peaks varies from about 30 cm$^{-1}$ to less than 10 cm$^{-1}$, where the sharpest peaks are too sharp to accurately determine the amplitudes.

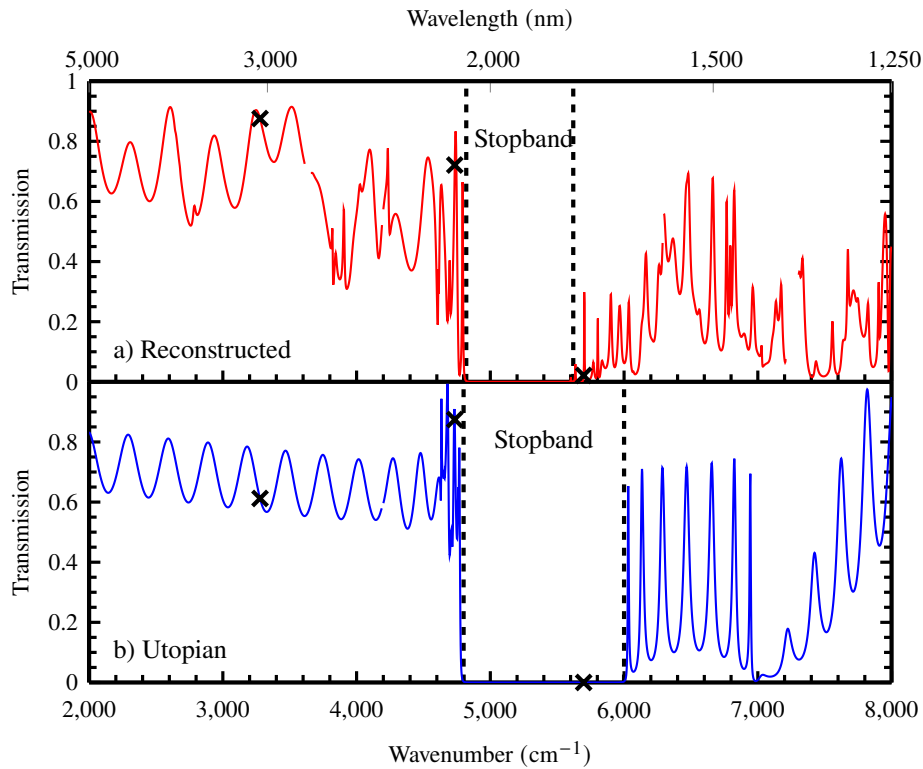

**Fig. 5.** Numerical transmission spectrum for (a) the reconstructed crystal (red curve), and (b) the utopian model crystal (blue curve), computed with $2\,\text{cm}^{-1}$ steps. The largest wavenumber range with transmission <1% is the stopband, and the boundaries are marked with dashed lines. The stopband for the reconstructed crystal is inside the stopband for the derived design crystal. The crosses mark the wavenumbers for which the fields are discussed in Section 4.3. The gaps in the spectrum are the result of omitting transmission values for the few wavenumbers that are inaccurate due to limitations of the PML, see Supplement 1.

(3) Above $5600\,\text{cm}^{-1}$ the transmission shows many peaks of varying width and amplitude that we attribute to random speckle. The minimum transmission is on average lower than in frequency range 1b (3600 to $4550\,\text{cm}^{-1}$).

### 4.2. Transmission of the utopian crystal

Figure 5(b) shows the transmission spectrum of the utopian crystal. Between 4792 and $6000\,\text{cm}^{-1}$ there is a stopband as the transmission is near zero ($T<1\%$). This stopband divides the frequency range into three parts, below, in and above the stopband.

(1a) Below the stopband up to about $4550\,\text{cm}^{-1}$ the spectrum consists of smooth fringes without sharp peaks, as expected for a perfectly periodic crystal. These Fabry-Pérot fringes have a spacing of $\Delta \nu = 270$ to $300\,\text{cm}^{-1}$ with an almost constant amplitude of about 20% and a slight downward trend. Comparing with the reconstructed crystal in the same spectral range, we see that the fringes match in spacing, but the fringes of the reconstructed crystal vary much more in amplitude. Moreover, unlike the reconstructed crystal, there is no transition to a spectral range (1b) with a mix of sharp peaks and fringes. Instead this spectral range is filled with the same regularly spaced fringes. This difference supports the hypothesis that the sharp peaks are speckle due to non-periodic structural features in the real crystal.

(1c) Near 4550 cm$^{-1}$ the spectrum transitions into a region where the transmission changes faster with wavenumber with four or more peaks between 4550 and 4792 cm$^{-1}$. These peaks result in a much larger variation in amplitude from 50% to 100% and a fast drop to near 0% when approaching the stopband at 4792 cm$^{-1}$. In the same spectral range (1c) of the reconstructed crystal, we see a similar pattern of sharp changes in transmission, but at lower amplitude. These features are tentatively assigned as band-edge features.

(2) Between 4792 and 6000 cm$^{-1}$ the transmission vanishes, hence there is a stopband. The stopband of the reconstructed crystal starts at 4812 cm$^{-1}$, within 0.5% of the real crystal. The close agreement of the low frequency edge of the stopband means that the averaging procedure of the utopian crystal is a faithful representation of the real crystal. Moreover it also confirms the common lore that the low frequency edge of a gap is robust to random disorder in a crystal [31]. A bigger difference is seen between 5600 and 6000 cm$^{-1}$ where the utopian crystal still has a stopband, but the reconstructed crystal shows gradually increasing transmission with some sharp peaks. The stopband of the reconstructed crystal is thus narrower with 15% relative bandwidth (ratio of band width and center frequency $\Delta\omega/\omega_c$) compared to the 22% for the utopian crystal.

(3) Above the stopband we distinguish two wavenumber regions with sharp peaks in transmission. Between 6000 and 7000 cm$^{-1}$ there are seven transmission peaks of about 70% spaced between 110 and 190 cm$^{-1}$ apart. At 7000 cm$^{-1}$ the transmission is low, probably due to a narrow higher-order stopgap in the bandstructure. Above 7000 cm$^{-1}$ there are four peaks visible spaced 190 cm$^{-1}$ with increasing maximum transmission to almost 100%. These predictable patterns in the transmission are completely different from the random speckle seen for the reconstructed crystal. Nevertheless, comparing the peaks between 6000 and 7000 cm$^{-1}$ there are transmission peaks at almost the same wavenumber, but with more varying maximum transmission.

The sharp peaks in the reconstructed crystal spectrum at frequencies inside the utopian stopband, so between 5600 to 6000 cm$^{-1}$, may be Anderson-localized that contribute to a Lifshitz tail in the density of states (DOS) in a gap, see also Refs. [71,72]. The unambiguous identification of Anderson-localized states requires extra steps such as finite-size scaling [73,74], which is outside the scope of our study.

### *4.3. Electric field distributions*

While the transmission contains information that is averaged over the whole thickness of the crystal, different physical quantities must be considered to obtain insights in the optical consequences of the structural features inside the crystal. To this end we now discuss the electric field distribution inside the crystal at several salient frequencies, which are marked with an 'x' in the transmission spectrum in Fig. 5.

Figure 6 shows the field at 3280 cm$^{-1}$, below the stopband in the fringe range (1a), where both crystals show clear Fabry-Pérot fringes without sharp peaks. The reconstructed and comparable design crystals have a transmission of 87% and 61% respectively, which is apparent in the different amplitude of the fringes in the air region. The field shows the outline of the pores, with the field inside the pores, and especially on the edges, being stronger than in the silicon. If we compare patterns and field strengths around different pores in the same crystal we observe a great similarity.

Figure 7 shows the fields at 4736 cm$^{-1}$, just below the stopband in the range of fringes and speckle (1c range). The transmission for the reconstructed and utopian crystal are almost the same at 81 % and 77 %. Nevertheless, there are significant differences between the field inside the real crystal and the utopian structure. In both crystals we observe that the field is strongest around the pores, but there is much more variation in the direction of propagation. For example, along the centerline of the crystal there are regions with almost zero field strength, and regions where the field is more than 2× the incident field strength. Comparing the real crystal with the utopian one we observe that the real crystal shows additional differences in the transverse direction, with the field on the bottom side being on average stronger than on the top.

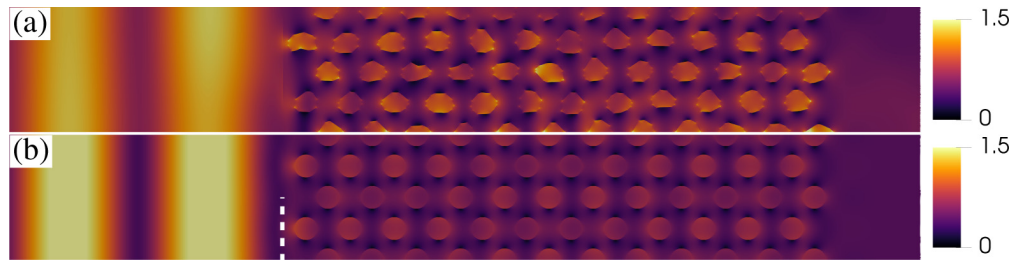

**Fig. 6.** Magnitude of the total electric field $|\mathbf{E}_{tot}|$ resulting from a plane wave at $\nu = 3280\,\text{cm}^{-1}$ incident on the (a) reconstructed and (b) utopian crystal with transmission of 87% and 61% respectively. The colorbar is normalized to 1.5 times the incident intensity. The dashed line in (b) indicates the position of the air-crystal interface for both crystals.

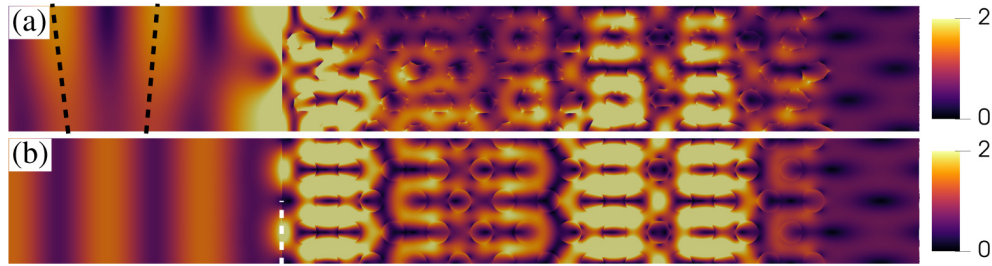

**Fig. 7.** Magnitude of the total electric field $|\mathbf{E}_{tot}|$ resulting from a plane wave at $\nu = 4736\,\text{cm}^{-1}$ incident on the (a) reconstructed and (b) utopian crystal with transmission of 81% and 77% respectively. The two dashed lines are guides to the eye to highlight that the wavefront is tilted with respect to the normal incident field. The colorbar is normalized to twice the incident intensity and is insufficient to resolve the locally very strong fields. The dashed line in (b) indicates the position of the air-crystal interface for both crystals.

At the front surface of the crystal we observe two differences between the reconstructed and utopian crystals. Firstly, the interference pattern of the incident and reflected wave has a different angle with respect to the interface. For the reconstructed crystal this is slightly tilted as highlighted by the dashed lines, while for the utopian crystal it is perfectly vertical. The diagonal interference pattern for the real crystal is the result of reflecting part of the light into the first mode. The utopian crystal and the incident beam are symmetric along the center line of the domain, reflecting the incident light into the first mode is incompatible with this symmetry. The varied pores in the real crystal break this symmetry, and thus allow reflecting into the first mode. Secondly, we note that there is a large difference in the field strength on the air side of the crystal. In specific cases such difference could be very important, for example when placing quantum dots on the surface. There are multiple factors influencing surface effects such as the distance between interface and pores and the shape of the closest pores, further analysis would be needed to determine the most important factor.

Figure 8 shows the fields at $5700\,\text{cm}^{-1}$. For the reconstructed crystal the transmission is 6 %, and the wavenumber is on the low-wavenumber side of a sharp transmission peak (that is not fully resolved by the computation, and whose maximum transmission is at least 30 %). For the utopian crystal the wavenumber is inside the stopband and the transmission is thus vanishingly small.

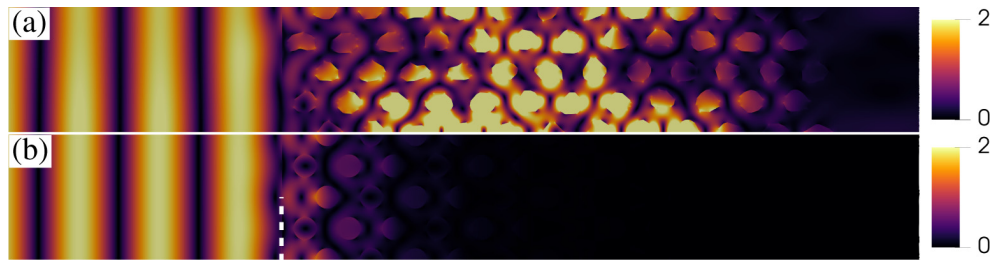

**Fig. 8.** Magnitude of the total electric field $|\mathbf{E}_{tot}|$ resulting from a plane wave at $\nu = 5700\,\text{cm}^{-1}$ incident on the (a) reconstructed and (b) utopian crystal with transmission of 6 % and 0 % respectively. The colorbar is normalized to twice the incident intensity and is insufficient to resolve the locally very strong fields. The dashed line in (b) indicates the position of the air-crystal interface for both crystals.

While the reflection in both cases is almost complete, the field inside the crystal is radically different: In the reconstructed crystal the field has an amplitude comparable to the amplitude of the incident field, with regions where the field has more than 2× the incident amplitude. The field is strongest in the bulk of the crystal away from the surfaces and reveals a random longitudinal and transverse distribution, somewhat similar to speckle in a (non-absorbing) random structure see, *e.g.*, Ref. [75]. For the utopian crystal we observe that the field decays exponentially away from the front interface of the crystal to values below the computational accuracy. The Bragg length corresponding to this decay is $l_{BR} = 775\,\text{nm}$, or about 1.6 lattice constants in the propagating direction. This is much less than the size of the crystal in propagation direction ($L = 6190\,\text{nm}$). Hence, since $l_{BR} \ll L$, the crystal is in the thick-crystal limit and the effects that we observe are bulk crystal effects with few surface effects expected.

To put this Bragg length in perspective with previous work: In Ref. [76] the Bragg length is derived from measurements of a real crystal with the same structure as studied here, but with slightly larger pores ($r/a = 0.21$, here 0.18). Based on the relative width of the stopband they compute a Bragg length of $l_{BR} = 2000\,\text{nm}$, a bit larger than here. However, if we use the same derivation from the bandwidth in our spectra we get a Bragg length of $l_{BR} = 2700\,\text{nm}$, a close agreement. The difference between the Bragg length derived from spatial intensity decay (its definition) versus the one derived from stopband width is attributed to the fact that in the latter a weak photonic interaction assumption is made [67], that is clearly violated in silicon-air nanostructures. In Ref. [62] the Bragg length was computed for a 3D inverse woodpile by varying crystal thickness. Such a crystal is not directly comparable due to the difference in dimension (3D vs 2D). Nevertheless, by converting their Bragg value to our crystal size, we get $l_{BR} = 180, 290\,\text{nm}$ for s and p-polarization, respectively, in reasonable mutual agreement.

## 5. Discussion

### 5.1. Utopian crystal as a model

In the vast majority of literature, utopian models are used to predict or understand the optical properties of real photonic crystals. By comparing the results of the real and the utopian crystal, we now address the question: "How reliable is the utopian model as a predictor or a source of explanations?"

We find that the frequency spectrum of the utopian model to be very ordered. Below the stopband there are Fabry-Pérot fringes, a short range of near-stopband effects, then a stopband, followed by ordered peaks above the stopband. Only part of this ordered structure is observed with the real crystal. Fabry-Pérot fringes are also present in the real crystal with accurately predicted spacing (within a few percent), but more varied amplitude. These fringes are only partially visible near the stopband, which we attributed to frequency speckle. The stopband and its lower band edge are accurately predicted (better than 0.5%), although the width is significantly smaller. The ordered peaks above the stopband of the utopian crystal completely miss the prediction of frequency speckle as with the real crystal.

In general we see that there is a better correspondence at low frequency than at high frequency. This is logical if we view the real crystal as a utopian crystal with additional scatterers from the deviating pores [22]. The frequency spectrum of the real crystal is then a combination of the ordered spectrum of the utopian crystal with additional features due to the inclusion of scatterers. The strength of these scatterers increases with frequency as their size relative to the wavelength increases. Correspondingly, the scattering mean free path due to all pore-variation scatterers decreases with frequency, increasing the effect of the scattering with respect to the background, and thereby decreasing the accuracy of the predictions from the utopian crystal.

With this model in mind we can also explain the great similarity of the electric fields at the lowest frequency. At such low frequency the pore-variation scatterers are very weakly scattering and the mean free path $l$ is much longer than the lattice size $a$. This is backed up by the presence of Fabry-Pérot fringes in the spectrum as for a stronger scattering sample with $l<L \approx 8.5a$ the scattering would disrupt an interference effect like Fabry-Pérot fringes.

Two of these robust features, the frequency of the bandgap and the spacing of the Fabry-Pérot peaks can be predicted using the crystal lattice and volume fraction, but without considering the exact pore shapes. For this we compute an effective refractive index [77] of the utopian unit cell by averaging the dielectric constant (70% silicon, 30% air) to obtain $n_{\text{eff}} = \sqrt{\langle \epsilon \rangle} \approx 2.96$. Approximating the crystal as a homogeneous block of length $L = 6190$ nm with this effective refractive index we obtain Fabry-Pérot fringes that are spaced at $\Delta \nu = (2Ln_{\text{eff}})^{-1} = 272$ cm$^{-1}$, while we observe 270 to 300 cm$^{-1}$ in excellent agreement. Similarly, by considering the crystal lattice and the corresponding Bragg peaks as in Ref. [23], we obtain an expected position of the stopband at $3/4 \times (cn_{\text{eff}})^{-1} = 5235$ cm$^{-1}$, practically in the middle of the stopband of the real crystal and on the lower half of the utopian crystal.

While the utopian model adds the additional accurate prediction on the lower edge of the stopband, it is fundamentally unable to predict the speckle observed both above and below the stopband. This speckle is the result of the varied shapes and size of the pores, ultimately resulting from entropic effects in the manufacturing. These features are lacking in the utopian model, and that is precisely what makes the utopian model utopian.

*5.2. Effects of boundary conditions*

The two mirror boundary conditions in the computational setup were used because of the limited width of the dataset. The use of mirror boundary conditions gives realistic pores on the edges of the domain, but has as side effect that the computational domain forms a waveguide. The use of such a waveguide geometry rather than a freestanding crystal has two implications.

The first implication is that at a fixed frequency only a few propagating modes are available. This is especially important in the air region, where the cutoff frequency of the $TE_1$ mode at 3649 cm$^{-1}$ is inside the relevant frequency range. Below this cutoff there is only a single propagating mode in the air region. Above this cutoff there are two propagating modes and we see two changes: there are no longer clear fringes in the transmission spectrum, and there is a reflection into the propagating $TE_1$ mode. Clear fringes, like those below the cutoff frequency, are expected when the wavelength is too long to be affected by the very local variation in pore shapes. The sudden transition in our results at the cutoff of the $TE_1$ mode is thus in all likelihood an artifact of our geometry. We expect that with a wider waveguide, and thus a lower cutoff frequency, the transition will happen at a lower frequency.

We note that the cutoff frequency for the $TE_1$ air-mode is the same for the utopian design. However, the utopian crystal has a width of two unit cells. The symmetry in the domain implies that the utopian crystal can only reflect and transmit into the $TE_0$, $TE_4$, $TE_8$, etc. air and silicon modes [78]. In the computed spectral range, the utopian crystal has one mode for reflection and up to two for transmission, while the reconstructed crystal has up to three modes for reflection and up to seven for transmission.

The second implication is that each mode on the reflection and transmission side corresponds to a specific frequency-dependent angle. The current computational setup thus allows to describe only an incident wave at these specific angles. The extension to true angle-resolved problems is theoretically simple. Applying the mirror boundary condition one gets periodic boundary conditions. Generalizing these to quasi-periodic would then support incident waves at arbitrary angles.

*5.3. Reconstruction*

An essential part of our study is the reconstruction process that was used to obtain the crystal structure from the X-ray data. Design choices such as representing the pore outline by a polygon, the algorithm used for the segmentation and the employed parameters all affect the shape and size of the pores in the reconstructed crystal, thereby affecting the transmission spectrum and electric fields presented here. However, the question is if the changes from a different design, algorithm or parameter will meaningfully influence the results and conclusions presented here. We discuss three different aspects of this question.

The effects of which segmentation algorithm is used and the subsequent choice for its parameters are naturally intertwined. Ideally, the effects are quantified through sensitivity analysis, but that is a laborious task. Instead we predict the effects based on the knowledge about the model. We note that changing the algorithm and parameter values changes the pores and thereby the resulting transmission spectrum. Thus, for the analysis we decompose the effect of changing pores into two parts, a change in pore size and a change in pore shape.

Firstly, we study the effect of the pore shapes. The reconstructed crystal consist of 60 pores each with a different shape. Thus, it is extremely unlikely that any similarities between the spectrum of the reconstructed and utopian crystals are accidental. It is far more likely, as we concluded, that these similarities are robust with respect to deviations in the pore shape.

Secondly, we consider the effect of varying the pore size. The effect of the pore size is taken into account by matching the pore size of the utopian crystal to that of the reconstructed crystal. Hence, the pore size cannot be the source of the differences between the reconstructed and utopian crystals. The strong influence on the similarities between the spectra of the reconstructed and utopian crystals is seen from the comparison with the effective refractive index model. The only free parameter in this model is the effective refractive index, which follows from the pore size via the silicon air ratio. The fact that this model accurately predicts the similarities, shows both that the pore size is important for the actual spectrum, and that it can predict its effects.

Thirdly, the reconstruction algorithm reconstructs the air-silicon interface as a polygon. Such a polygon has corners, and these are known to create locally very strong fields from corner singularities [79]. Indeed, in the computed electric fields we observe locally very strong fields at the corners of the pores. The X-ray data does not have sufficient resolution to determine whether such corners exist, and we do not necessarily expect them from the manufacturing process. Therefore, it is not unlikely that the air-silicon boundary is smoother in the original crystal than in the reconstructed crystal. Such a smoother boundary could be made by describing the interface by curved line segments. The usefulness of doing this depends on what the results are used for. Curved boundaries will remove or reduce the corner singularities and locally reduce the very extreme fields. We expect that the effect on global properties like the transmission is limited, but that it is more important for local quantities like the LDOS.

*5.4. Outlook: real crystal computation*

The computation presented here is a first demonstration of a data-driven photonic crystal computation. As already stated in the introduction of this new approach, we see a broad application to all kinds of photonic nanostructures including waveguides, meta surfaces, plasmonic structures and topological photonical materials. For our specific application to photonic crystals we envision three broad directions how our method can be extended.

Firstly, the computations presented here consider a plane wave incident on a 2D waveguide setup, which is unlike the experimental setup of focussing a beam on a 3D finite size crystal. A better understanding of the experimental situation will be gained by changing the computational setup to more closely align with reality. Such extensions are essential in understanding real samples, as the current 2D waveguide setup prohibits certain observed effects like mixing between polarizations. However, an extension to a computation of a completely realistic 3D situation is far beyond current computational capabilities. Therefore, additional assumptions, like 2D or finite width, will be necessary to reduce the complete measured crystal to a model that is computationally tractable. These assumptions influence the numerical results preventing direct comparison with experimental results. Nevertheless, by quantifying these influences a limited but meaningful comparison is feasible between on the one hand experimental results and on the other hand computations that invoke a realistic but simplified model based on the measured structure of the same crystal. We expect that such comparison is currently feasible.

A second direction is to consider several forms of sensitivity analysis based on the current computational setup. The sensitivity of the results with respect to the parameters used in the reconstruction will give a more rigorous idea on how accurate we need to be in the reconstruction. This could be used to verify our expectation that the spectral features that are shared by the reconstructed and utopian crystals are relatively robust. Moreover, looking at sensitivity to local deformations of the pore shapes may provide better insight into whether there are types of deviations that more strongly influence the performance of the crystal structure and thus require attention in the manufacturing process. Such a sensitivity could be obtained by reconstructing and computing several parts or slices of a larger dataset and mutually comparing them, which is a major undertaking. This would give a qualitative impression of the changes in the optical properties in different parts of the crystal.

A third direction is to focus on how accurate the reconstruction of the real crystal is, with the goal of creating the perfect reconstruction for the best comparison with experimental work. The current approach to the reconstruction and especially the selection of the parameters is based on visual judgment of the result. One consequence of this is that the pore size is likely different from that of the actual crystal. Such differences could be resolved by calibrating the reconstruction process against the actual structure. Such a calibrated reconstruction will provide a better candidate for a comparison between computation and actual measurements, and therefore an understanding of the individual sample.

## 6. Conclusions

In conclusion, we have developed and presented a new data-driven approach to computational nanophotonics. The essence of this approach is to use a reconstruction of a real nanostructure, including all its manufacturing deviations. These deviations significantly impact the optical performance of real nanostructures, but are absent from traditional computational nanophotonics where one considers structures so perfect that they can only be made in a manufacturing Utopia. Comparison of traditional computational results with real measurements is thus fraught with many possible causes of differences, but there is no way to distinguish between them. Our method provides an intermediate between these two extremes that allows better understanding in the relevance of theoretical results and insight in the working of actual devices.

To demonstrate our new method, we apply it to a X-ray imaging reconstruction of a photonic bandgap crystal with inverse woodpile structure. This is done by processing a 2D slice from the reconstruction in a computational mesh. This mesh is then used to compute the transmission through the reconstructed crystal over a broad frequency range. For a comparison with existing models we consider the transmission through a matching utopian crystal, a crystal with perfectly round pores and matching fill fraction.

For the reconstructed crystal we observe Fabry-Pérot fringes, a stopband and frequency speckle. The utopian model shows similar Fabry-Pérot fringes and a slightly wider stopband. The speckle seen with the reconstructed crystal is completely absent from the utopian model, and the transmission above the stopband is thus completely different.

The use of computational methods allows, unlike experiments, to study the fields everywhere inside the crystal in detail. We have thus compared the fields inside the reconstructed and utopian crystal at three sample frequencies. We observed major differences in the fields at constant frequency, which are not apparent from external probes like transmission.

**Funding.** Nederlandse Organisatie voor Wetenschappelijk Onderzoek (JCSER 680-91-084, FOM 138, NWO-TTW P15-36).

**Acknowledgments.** We thank Peter Cloetens, Melissa Goodwin, Diana Grishina, Cock Harteveld, Jens Wehner for help. We thank ESRF for beamtime through experiments HC-2520 and CH-5092.

**Disclosures.** The authors declare no conflicts of interest.

**Data availability.** Data underlying the results presented in this paper are available in Ref. [53]. The original 3D dataset is not publicly available at this time but may be obtained from the authors upon reasonable request.

**Supplemental document.** See Supplement 1 for supporting content.

# Non-utopian optical properties computed of a tomographically reconstructed real photonic nanostructure: supplement

LARS J. CORBIJN VAN WILLENSWAARD,[1,2] STEF SMEETS,[3] NICOLAS RENAUD,[3] MATTHIAS SCHLOTTBOM,[2] JAAP J. W. VAN DER VEGT,[2] AND WILLEM L. VOS[1,*]

[1]*Complex Photonic Systems (COPS), MESA+ Institute for Nanotechnology, University of Twente, P.O. Box 217, 7500AE Enschede, The Netherlands*
[2]*Mathematics of Computational Science (MACS), MESA+ Institute for Nanotechnology, University of Twente, P.O. Box 217, 7500AE Enschede, The Netherlands*
[3]*Netherlands eScience Center, Science Park 402, 1098 XH Amsterdam, The Netherlands*
*https://nano-cops.com*
*[*]w.l.vos@utwente.nl*

# Non-utopian optical properties computed of a tomographically reconstructed real photonic nanostructure: supplemental document

## 1. TRANSVERSE MODES

The two mirror boundaries on the top and bottom of the domain create a waveguide like geometry. This creates a problem with the PML, which can not absorb these modes near their cutoff wavenumber. Here we highlight the cause of this problem.

The mirror boundaries at $x = 0$ and $x = L_x$ restrict the momentum in $x$-direction to

$$k_{x,m} = \frac{m\pi}{L_x \sqrt{\epsilon \mu}}, \tag{S1}$$

where $m$ is an integer and $\epsilon$ and $\mu$ are the relative permittivity and permeability [1]. The corresponding modes have a field that is the interference pattern of two plane waves

$$\mathbf{E}_m(x,y) = e^{ik_y y} \left[ \begin{pmatrix} 1 \\ -\frac{k_{x,m}}{k_y} \end{pmatrix} e^{ik_{x,m}x} + \begin{pmatrix} 1 \\ \frac{k_{x,m}}{k_y} \end{pmatrix} e^{-ik_{x,m}x} \right] = e^{ik_y y} \begin{pmatrix} \cos(k_{x,m}x) \\ -2\mathrm{i}\frac{k_{x,m}}{k_y} \sin(k_{x,m}x) \end{pmatrix} \tag{S2}$$

with the momentum in $y$-direction following from the wavenumber $\nu$ and the dispersion relation

$$\mu \epsilon \nu^2 = 4\pi^2 (k_{x,m}^2 + k_y^2). \tag{S3}$$

Each mode has a cutoff wavenumber $\nu_c = 2\pi k_{x,m}/\sqrt{\epsilon \mu}$ where the mode transitions from evanescent behavior with imaginary $k_y$ to propagating behavior with real $k_y$.

The spurious reflection of these modes from the PML depends on the design of the PML and whether a mode is evanescent or propagating. For both propagating and evanescent modes the spurious reflection $R$ follows the relation [2]

$$\ln(R) \propto -|k_y|. \tag{S4}$$

This is a problem near the cutoff wavenumber, because $|k_y| \to 0$ and $R \to 1$ as $\nu$ approaches the cutoff. The spurious reflection will thus approach 1 near the cutoff wavenumber. The properties of the PML, like the profile and thickness, can adjust the rate at which it approaches 1. Adjusting these will impact the computational cost and the spurious reflection from the discretization.

For the computations presented here, we choose to adjust these parameters to reduce the spurious reflection in the neighbourhood of the cutoff frequencies. For the frequencies very close to the cutoff the parameters are not sufficient to completely remove the spurious reflection. Therefore, we inspect the amplitude of the solution at the far end of the PML. If the amplitude is larger than 0.1 times the incident amplitude, then the spurious reflection is at least 1% and we discard the transmission value for that wavenumber as inaccurate.

## 2. LOG TRANSMISSION

Figure S1 shows the transmission spectrum of Figure 5 with log-transmission scale. We observe that the transmission decreases rapidly at the upper and lower edges of the stopband. Inside the stopband the transmission is smaller than the violation of energy conservation $e = |T + R - 1|$. This energy violation is a lower bound for the numerical error in the transmission, therefore transmission values less than $e$ are considered to be not significant and have thus been omitted.

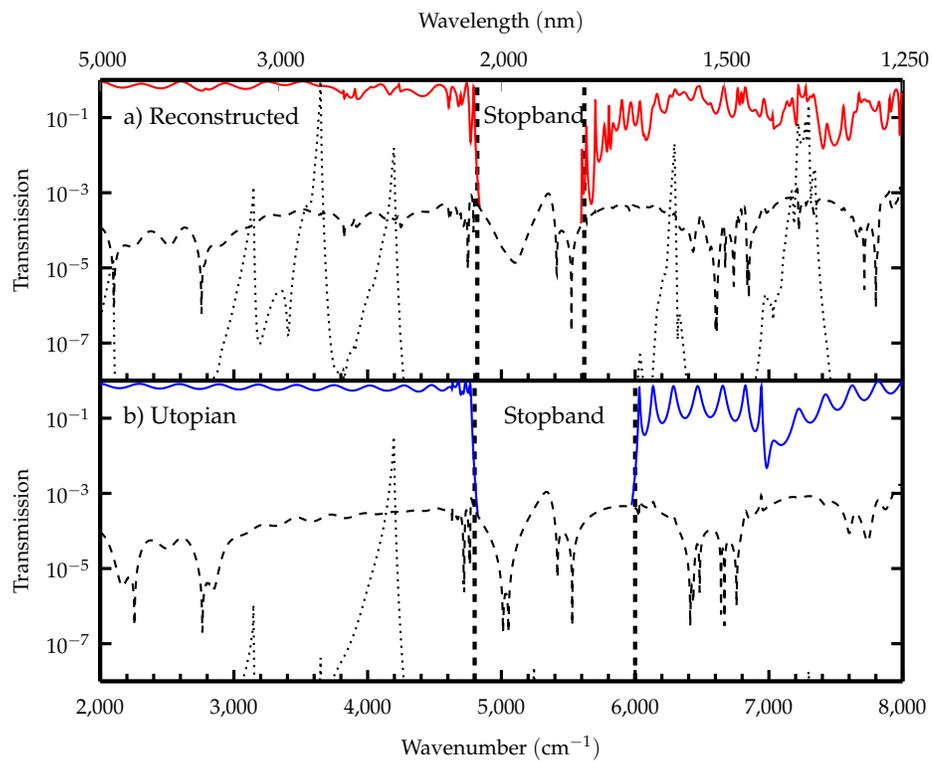

**Fig. S1.** Numerical transmission spectrum from Figure 5 with transmission on a log scale, (a) the reconstructed crystal and (b) the utopian model crystal. The black lines correspond to two sources of numerical errors, (dashed) the error in energy conservation $e = |T + R - 1|$ and (dotted) spurious reflection of the PMLs (see Appendix 1). Transmission values that are below either of these curves are left out.

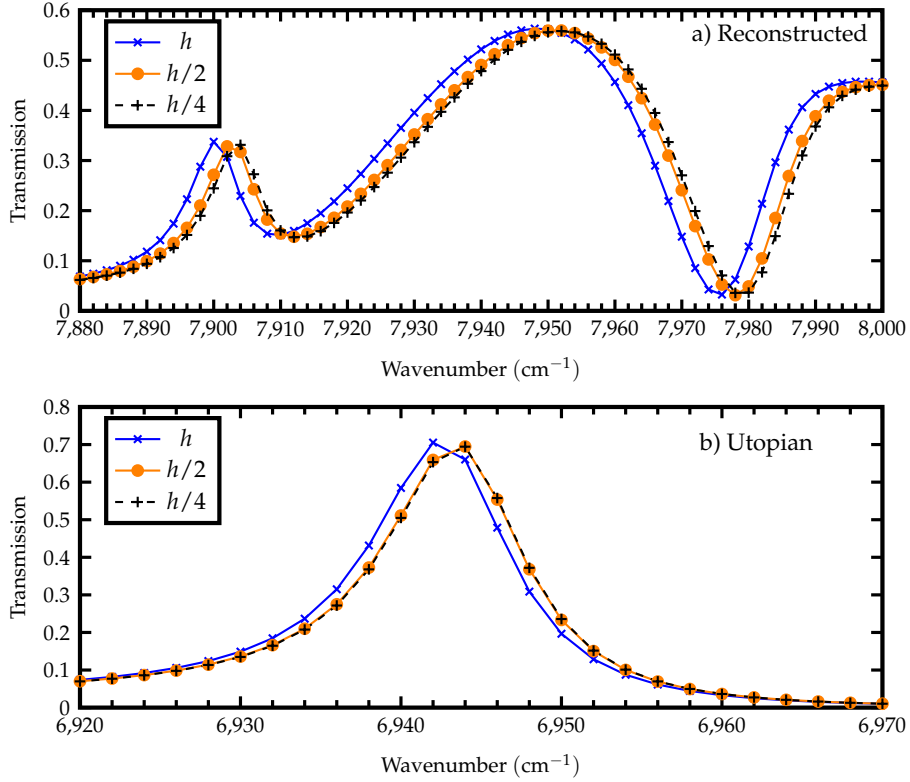

**Fig. S2.** Small section from the numerical transmission spectrum for the a) reconstructed crystal computed on three meshes, approximately doubling the mesh resolution in each step. Each mark corresponds to one computation, connected by straight lines to guide the eye. The transmission curve at higher mesh resolutions have the same shape, but the wavenumber axis is stretched. The same effect is visible for the b) utopian crystal, with the transmission for the smallest two meshes practically overlapping. The location of both sections is chosen to make the differences between the three meshes clearly visible.

## 3. NUMERICAL ACCURACY

To investigate the accuracy of our results we study the effect that mesh refinement has on the transmission spectrum. For both crystals we consider meshes of three different resolutions. For the utopian crystal we start with a coarse mesh with elements of size $h$. Each curvilinear triangle is subdivided into four smaller ones to obtain a mesh with size $h/2$, and a second time to get a mesh of size $h/4$. This subdivision was not possible with the mesh generated for the reconstructed crystal, hence we created three different meshes. The mesh generated for size $h$ specifies for each part of the domain (PML, crystal, etc.) the maximum area of an element, resulting in a mesh with 18351 triangles. For meshes of size $h/2$ and $h/4$ the constraints on the maximum area was divided by $2^2$ and $4^2$ respectively resulting in meshes with 72006 and 287665 triangles respectively. The coarsest mesh (size $h$) for the reconstructed mesh is shown in Figure 3c, the transmission spectra and fields presented in the main text are using the meshes with resolution $h/4$.

We computed the transmission spectrum for all three meshes of both crystals, keeping all other parameters identical. Figure S2a shows a small part of the transmission spectrum for the reconstructed crystal computed on these three meshes. We observe that shape of the three transmission spectra are visually identical, but they are offset with respect to each other. This shift is caused by the numerical method introducing a slight error in the phase, thereby stretching or compressing the wave [3]. This effect increases with wavenumber and larger elements. Compared to the best resolution ($h/4$), we see a shift of slightly more than 2 and less than 1 wavenumber step for the meshes of size $h$ and $h/2$ respectively. This shows that the wavenumber error of our results is smaller than 1 wavenumber step of $2\,\text{cm}^{-1}$.

Figure S2b shows a similar transmission plot for the utopian crystal. This is zoomed on a very sharp peak at slightly lower wavenumber. We observe almost no difference between the transmission spectra of the meshes with resolution $h/2$ and $h/4$, the transmission spectrum for the mesh with resolution $h$ is shifted by approximately 1 wavenumber step corresponding to $2\,\text{cm}^{-1}$.

In conclusion, the finite mesh size introduces a wavenumber error. For the meshes used (size $h/4$) the error is smaller than 1 wavenumber step of $2\,\text{cm}^{-1}$ for the reconstructed crystal and negligible for the utopian crystals. Hence, the effects of a finite mesh are so small that they do not impact the conclusions presented in the main text.

## 4. NUMERICAL METHOD

To compute the transmittance we split the total electric field $\mathbf{E}_{\text{tot}} = \mathbf{E}_s + \mathbf{E}_b$ into a known background field $\mathbf{E}_b$ and numerically computed scattered field $\mathbf{E}_s$. This scattered field is the solution to the following variant of the time harmonic Maxwell's equations

$$\nabla \times \mu^{-1} \nabla \times \mathbf{E}_s - \nu^2 \epsilon \mathbf{E}_s = \nabla \times \mu^{-1} \nabla \times \mathbf{E}_b - \nu^2 \epsilon \mathbf{E}_b, \tag{S5}$$

where $\mu$ and $\epsilon$ are the relative permeability and permittivity, respectively, $\nu$ is the wavenumber. For simplicity we will denote the known right hand side by $\mathbf{j}$. The use of a PML requires the modification of the permeability and permittivity in the PML regions to tensors ($\overleftrightarrow{\mu}$ and $\overleftrightarrow{\epsilon}$) and assumes $\mathbf{j} = \mathbf{0}$ in these regions. We therefore solve

$$\nabla \times \overleftrightarrow{\mu}^{-1} \nabla \times \mathbf{E}_s - \nu^2 \overleftrightarrow{\epsilon} \mathbf{E}_s = \mathbf{j}, \tag{S6}$$

and assume perfectly conducting boundary conditions $\mathbf{n} \times \mathbf{E}_s = \mathbf{0}$ with $\mathbf{n}$ the outward normal.

### A. Finite element method

We use an interior penalty discontinuous Galerkin finite element method (IPDGFEM) to solve Eq. S6 numerically. Versions of this method, with additional simplifying assumptions for the theoretical analysis, have been proposed and analyzed in literature [4–7].

Let $\mathcal{T}_h$ be a triangulation of the computational domain $\Omega$, and denote the set of faces by $\mathcal{F}_h$. As solution space we use the space spanned by the discontinuous second order Nédélec functions of the first kind on this triangulation [8]. Specifically, given an triangle $T$ define the Nédélec functions as

$$\mathcal{E}(T) = \left\{ \mathbf{v} \in [\mathcal{P}_2(T)]^2 \,\Big|\, \mathbf{v} \cdot \mathbf{t} \in \mathcal{P}_2(e_j),\, j = 1, 2, 3 \right\}, \tag{S7}$$

where $\mathcal{P}_2(D)$ are the complex valued polynomials of degree at most 2 on a domain $D$, and $e_j$ the edges of the triangle with $\mathbf{t}$ as the tangent vector. The finite element space is formed by

$$V = \left\{ \mathbf{v} \in \left[L^2(\Omega)\right]^2 \,\Big|\, \mathbf{v}|_T \in \mathcal{E}(T),\, \forall T \in \mathcal{T}_h \right\}, \tag{S8}$$

where $L^2(\Omega)$ are the square integrable functions on the computational domain $\Omega$. The extension to curvilinear elements, which is used to more accurately represent the circular pores of the utopian crystal, follows standard procedure of using isoparametric maps [8].

A function $\mathbf{v} \in V$ is two valued on any internal face $F$ between elements $T_1$ and $T_2$. Denote by $\mathbf{v}_1$ and $\mathbf{v}_2$ the values as limit from $T_1$ and $T_2$, respectively, and denote by $\mathbf{n}_1$ and $\mathbf{n}_2$ the unit outward normal on the face from $T_1$ and $T_2$, respectively. On a face we define the following jump and average operators:

$$[\![\mathbf{v}]\!]_T = \mathbf{n}_1 \times \mathbf{v}_1 + \mathbf{n}_2 \times \mathbf{v}_2 \qquad \text{and} \qquad \{\!\!\{\mathbf{v}\}\!\!\} = \frac{1}{2}(\mathbf{v}_1 + \mathbf{v}_2). \tag{S9}$$

On a boundary face $F$ with outward normal $\mathbf{n}$ we define them as

$$[\![\mathbf{v}]\!]_T = \mathbf{n} \times \mathbf{v} \qquad \text{and} \qquad \{\!\!\{\mathbf{v}\}\!\!\} = \mathbf{v}. \tag{S10}$$

The sesquilinear form for IPDGFEM is defined as

$$\begin{aligned}
a_h(\mathbf{E}_s, \mathbf{v}) = & \sum_{T \in \mathcal{T}_h} \int_T \left( \overleftrightarrow{\mu}^{-1} \nabla \times \mathbf{E}_s \right) \cdot \overline{\nabla \times \mathbf{v}} \, dx \\
& - \sum_{F \in \mathcal{F}_h} \int_F [\![\mathbf{E}_s]\!]_T \cdot \{\!\!\{ \overleftrightarrow{\mu}^{-H} \overline{\nabla_h \times \mathbf{v}} \}\!\!\} + \{\!\!\{ \overleftrightarrow{\mu}^{-1} \nabla_h \times \mathbf{E}_s \}\!\!\} \cdot \overline{[\![\mathbf{v}]\!]_T} \, dS \\
& + \sum_{F \in \mathcal{F}_h} \int_F \frac{a_F}{h_F} [\![\mathbf{E}_s]\!]_T \cdot \overline{[\![\mathbf{v}]\!]_T} \, dS,
\end{aligned} \tag{S11}$$

where $\bar{\mathbf{v}}$ denotes the complex conjugate of $\mathbf{v}$, $a_F$ is a stabilization parameter, that we set to 100, $h_F$ is the diameter (length) of face $F$, and $\mu^{-H}$ is the inverse of the Hermitian transposed of the tensor $\mu$.

The numerical problem is to approximate $\mathbf{E}_s$ by a function from $V$ that satisfies

$$a_h(\mathbf{E}_s, \mathbf{v}) - \nu^2 \int_\Omega \left(\overleftrightarrow{\epsilon} \mathbf{E}_s\right) \cdot \bar{\mathbf{v}} \, dx = \int_\Omega \mathbf{j} \cdot \bar{\mathbf{v}} \, dx \qquad \forall \mathbf{v} \in V. \tag{S12}$$

The resulting system of linear equations is solved with a direct solver.

## B. PML

The implementation of the PML directly follows [3]. Specifically, consider a PML region that should absorb waves in $y$ direction from $y = 0$ to $y = L_{\text{pml}}$ from a region with material coefficients $\epsilon$ and $\mu$. Define

$$d(y) = 1 + \frac{i}{\nu} a y^2, \tag{S13}$$

where i is the imaginary unit and $a$ is a parameter that defines the strength of the damping. The material tensors in the PML layer are defined as

$$\overleftrightarrow{\epsilon}(y) = \epsilon \, \text{diag}(d(y), 1/d(y)), \tag{S14a}$$

$$\overleftrightarrow{\mu}(y) = \mu \, \text{diag}(d(y), 1/d(y)), \tag{S14b}$$

where $\epsilon$ and $\mu$ are the material parameters of the underlying material, e.g. the ones for silicon for a PML attached to a silicon region.

The parameter $a$ is computed so that the PML dampens a normal incident plane wave by a factor $R_0 = 10^{-40}$ via the formula

$$\log R_0 = -\frac{2}{3} \sqrt{\epsilon} a L^{-3}, \tag{S15}$$

where $L$ is the depth of the PML. The very small spurious reflection, $R_0 = 10^{-40}$, only applies to the lowest order mode, for higher order modes the spurious reflection is wavenumber dependent. The choice $R_0 = 10^{-40}$ ensures that spurious reflection of higher order modes decays quickly above their cut-off wavenumber. This can be seen in dotted lines in Figure S1, where the spurious reflection drops quickly to the right of each peak.